\documentclass[prd,english,preprintnumbers,amsmath,amssymb,nofootinbib,twocolumn,superscriptaddress]{revtex4-1}

\pdfoutput=1

\usepackage[latin1]{inputenc}
\usepackage{graphicx}
\usepackage{bbm}
\usepackage{amssymb}
\usepackage{amsmath}
\usepackage{tabularx}
\usepackage{bigints}
\usepackage{hyperref}
\hypersetup{colorlinks=false}

\usepackage{color}

\usepackage{dsfont}

\def\0#1#2{\frac{#1}{#2}}

\def\s0#1#2{\mbox{\small{$ \frac{#1}{#2} $}}}

\newcommand{\Tr}{\mathrm{Tr}}

\newcommand{\m}[1]{\ensuremath{\mathrm{#1}}}
\newcommand{\dif}[1]{\ensuremath{\medspace \mbox{d} #1}}

\newcommand{\gdif}[2]{\ensuremath{\frac{\dif{#1}}{\dif{#2}}}}

\usepackage{babel}
\makeatother
\begin{document}

\title{Critical $O(2)$ field theory near six dimensions beyond one loop}

\author{Dietrich Roscher}
\affiliation{Department of Physics, Simon Fraser University, Burnaby, British Columbia, Canada V5A 1S6}
\affiliation{Institute for Theoretical Physics, University of Cologne, 50937 Cologne, Germany}
\author{Igor F. Herbut}
\affiliation{Department of Physics, Simon Fraser University, Burnaby, British Columbia, Canada V5A 1S6}

\begin{abstract}
A tensorial representation of $\phi^4$  field theory introduced in Phys. Rev. D. 93, 085005 (2016) is studied close to six dimensions, with an eye towards a possible realization of an interacting conformal field theory in five dimensions. We employ the  two-loop $\epsilon$-expansion, two-loop fixed-dimension renormalization group, and non-perturbative functional renormalization group. An interacting, real, infrared-stable fixed point is found near six dimensions, and the corresponding anomalous dimensions are computed to the second order in small parameter $\epsilon=6-d$. Two-loop epsilon-expansion indicates, however,  that the second-order corrections may destabilize the fixed point at some critical $\epsilon_c <1$. A more detailed analysis within all three computational schemes suggests that the interacting, infrared-stable fixed point found previously collides with another fixed point and becomes complex when the dimension is lowered from six towards five. Such a result would conform to the expectation of triviality of $O(2)$ field theories above four dimensions.
\end{abstract}

\maketitle

\section{Introduction}

The question of existence of conformally  invariant interacting field theories in (space-time) dimensions higher than four has recently stimulated efforts in two closely related directions. In the first~\cite{Fei2014,Fei2015}, it was shown that the $O(N)$-symmetric  $\phi^4$-theory at negative self-interaction can be Hubbard-Stratonovich-decoupled in the scalar channel, and then written in a form that admits a non-Gaussian, real, infrared (IR) stable interacting fixed point close to and below six dimensions, for a sufficiently large parameter $N$.  In the related development~\cite{OneLoop16}, it was pointed out that there exists an alternative decoupling of the same theory in the tensor channel, in which the $O(N)$ theory would be written as
\begin{equation}
\label{BaseModel}
\mathcal{L} = \frac{1}{2}(\partial_\mu z_a)^2 + \frac{1}{2}(\partial_\mu\phi_i)^2 + g z_a\phi_i\Lambda^a_{ij}\phi_j + \lambda\m{Tr}\left[(z_a\Lambda_a)^3\right].
\end{equation}
Here $i, j = 1,2,...N$,  $\phi_i$ and $z_a$ are real fields, with $a=1, ... M_N$, and $M_N = (N-1)(N+2)/2$ is the number of components of the irreducible tensor of the second rank under $O(N)$ rotations. The $M_N$ matrices $\Lambda^a_{ij}$ provide a basis in the space of traceless, real, symmetric $N$-dimensional matrices, and the fields $z_a$  transform as components of a second-rank tensor. The theory (1) reproduces the original formulation~\cite{Fei2014} if one would retain a single matrix $\Lambda$ and replace it with a unit matrix. The two IR-relevant mass terms for the fields $\phi_i$ and $z_a$  have been tuned to zero for simplicity.  A perturbative one-loop analysis~\cite{OneLoop16} of (1) in dimension $d=6-\epsilon$ identified a non-trivial real IR-stable fixed point in certain ranges of small values of $N$ that include, most interestingly, the physically relevant cases of $N=2$ and $N=3$. Such an indication of the possibility of a successful UV completion of $O(N)$ theory above four dimensions, on the other hand, would appear to be somewhat at odds with expectations based on well-known earlier results on the subject; in~\cite{Aizenman81,FROHLICH1982} it was proven, for example, that the standard $\phi^4$ theory in (integer) dimensions $ d > 4$ is bound to be trivial for $N=1$ and $N=2$.

The structural equivalence between the standard $\phi^4$ theory and our model can be established by integrating out the $z_a$ fields in eq.~\eqref{BaseModel}. Doing so yields a negative quartic coupling~\cite{OneLoop16}, with some residual momentum dependence, so it seems unclear whether the above mentioned no-go theorems should indeed apply to the tensorial representation of the $O(N)$ theory. At any rate, we take this conceptual tension as an additional motivation for further studies of  the field theory (1). In this work we therefore reconsider the eq. (1) for $N=2$ since it represents the most convenient departure point for any deeper analysis due to its particular simplicity: the matrices $\Lambda_{ij}^a$ in this case reduce to the two Pauli matrices $\sigma_x$ and $\sigma_z$. Since any trace over a product of three of these matrices vanishes, so does the term cubic in the $z_a$ fields in eq.~\eqref{BaseModel}. This drastically reduces the computational cost of higher order (perturbative) investigations.

As a first step, we extend the $\epsilon$-expansion performed in~\cite{OneLoop16} to the two-loop level. While the non-trivial fixed point of the beta function  for the coupling $g$ is of course a power series and thus continuous in the parameter $\epsilon=6-d$, the $\epsilon^2$ correction to the leading result destabilizes it at a certain critical value $\epsilon <1$. Examining the non-trivial roots of  the beta-function, one finds that the fixed point is real only for $\epsilon\ll 1$, being rendered complex on its way to the physical $\epsilon=1$ by the collision and annihilation with another fixed point. This is less surprising after recalling that the same occurs at some critical $\epsilon$ at every even order of expansion around $d=4$ in the canonical $\phi^4$ theory as well~\cite{Kleinert2001}. It is conceivable  that in our theory this conclusion would change from one order to the next, if no summation of the series is performed. Nevertheless, since unlike in the standard $\phi^4 $ case, here we do not {\it a priori} know that the critical point in $d=5$ exists, we explicitly entertain the possibility that it may  disappear on its way from six to five dimensions. The reliability of any conclusion in $d=5$ is of course diminished by the fact that there seem to be no signs of  convergence of the obtained series in powers of $\epsilon$. This motivates us to perform two additional, independent calculations in different, yet comparable renormalization group (RG) schemes.

Directly at $d=5$, a technique known as \emph{fixed dimension renormalization group } (RG) can be employed. Introduced by Parisi~\cite{Parisi1998}, it has provided very precise estimates of critical exponents  for standard universality classes in three dimensions~\cite{Kleinert2001}. Aside from minor quantitative variations, the results in this scheme agree qualitatively with those of the $\epsilon$-expansion: to two-loop order, besides the Gaussian, we only find complex zeros of the beta-function, and thus no real non-trivial fixed point.

The destruction of the stable fixed point that exists near six dimensions in the perturbative treatment occurs due to a collision with another, non-Gaussian fixed point. The value of the latter is not $\mathcal{O}(\epsilon)$ and therefore, although being accidentally small, strictly speaking not in a regime where perturbative methods should be trusted. Our third approach is therefore to utilize functional RG (fRG), an inherently non-perturbative technique. While we can indeed confirm the destruction of the fixed point at some intermediate dimension $5<d<6$, a comparison with the first two calculations is not as straightforward as between the latter, and calls for a closer examination in the future.

This paper is organized as follows. In Sec.~\ref{SecEps}, we extend the $\epsilon$-expansion to two-loop level and analyze the ensuing RG flow equations. Sec.~\ref{SecFixDim} is devoted to the application of fixed dimension RG, while Sec.~\ref{SecfRG} discusses the functional RG approach. In Sec.~\ref{SecConc}, we offer some concluding remarks.

\section{Tensorial $O(2)$ model and $\epsilon$-expansion}
\label{SecEps}
\begin{figure}[t]
\centering

\includegraphics[width=0.5\textwidth]{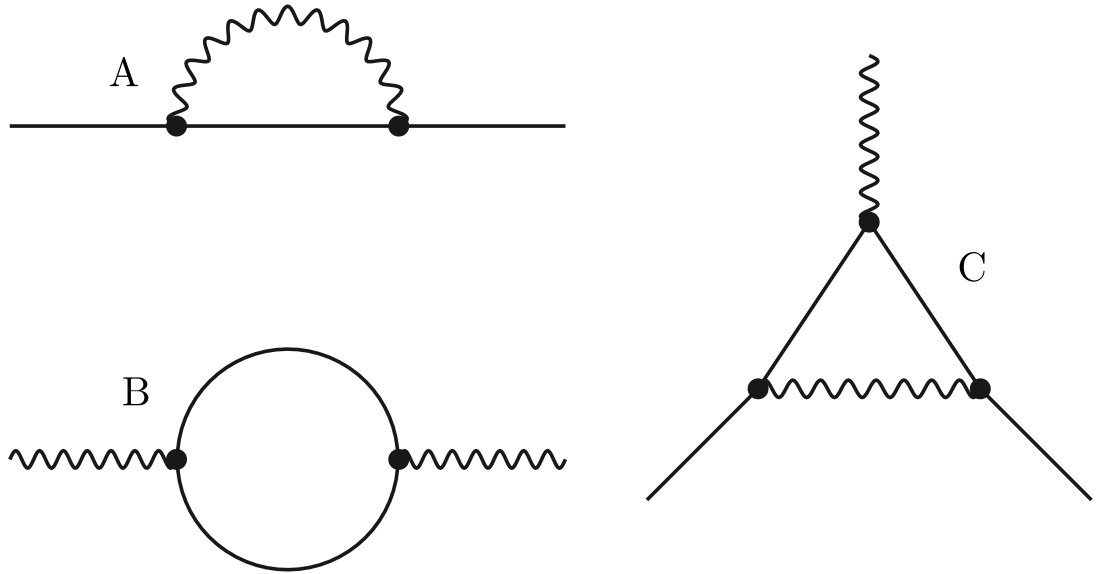}
\caption{One loop diagrams contributing in principle to the RG flows of $\eta_\phi$ (A), $\eta_z$ (B) and $d g_R/d\ln b$ (C). The solid lines correspond to $\phi$ (scalar field) propagators whereas the wiggly lines symbolize $z^a$ (tensor field) propagators. The diagram C vanishes.}
\label{OneLoopDiags}
\end{figure}

For the remainder of this work we confine ourselves to $N=2$. As argued in the introduction, no term cubic in the $z_a$ fields is present in this case and eq.~\eqref{BaseModel} is reduced to
\begin{equation}
\label{O2Model}
\mathcal{L} = \frac{1}{2}(\partial_\mu z_a)^2 + \frac{1}{2}(\partial_\mu\phi_i)^2 + g z_a\phi_i\sigma^a_{ij}\phi_j,
\end{equation}
where $\sigma^a\in\{\sigma_x,\sigma_z\}$. The corresponding one-loop RG equations have already been computed in~\cite{OneLoop16}. In Fig.~\ref{OneLoopDiags} we give the corresponding Feynman diagrams to be evaluated for this purpose.
It is particularly noteworthy that the correction to the vertex (Fig.~\ref{OneLoopDiags}C) vanishes due to the peculiarities of the Pauli algebra. This drastically reduces the number of diagrams contributing at two-loop level as well. Every diagram that contains Fig.~\ref{OneLoopDiags}C as a substructure is bound to vanish. Furthermore, two-loop vertex correction diagrams which are built from Fig.~\ref{OneLoopDiags}C and propagator correction diagrams such as Figs.~\ref{OneLoopDiags}A and \ref{OneLoopDiags}B effectively still exhibit the same Pauli structure and therefore vanish as well.

\begin{figure}[t]
\centering
\includegraphics[width=0.5\textwidth]{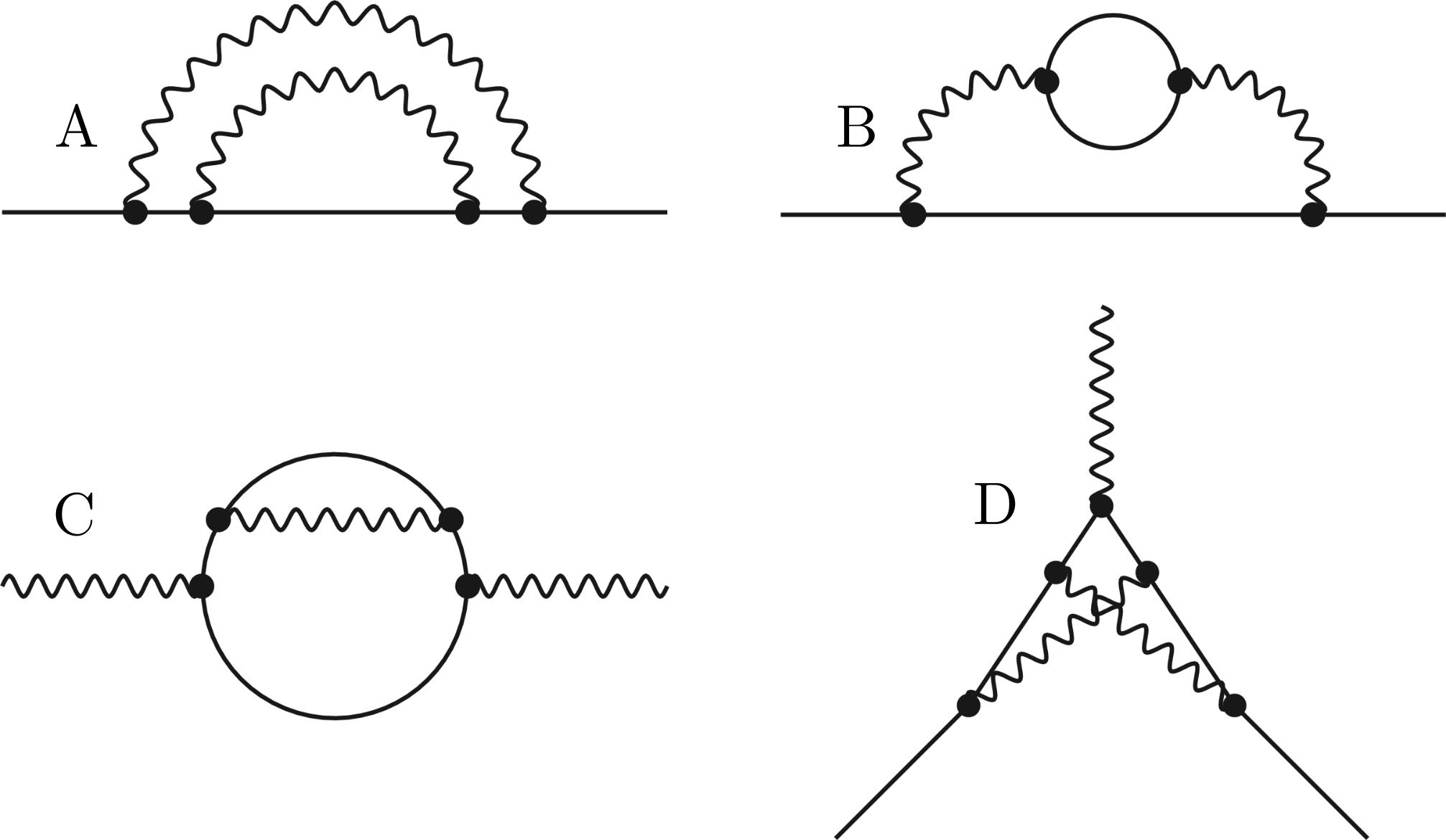}
\caption{Non-vanishing two loop diagrams contributing to the RG flows of $\eta_\phi$ (A,B), $\eta_z$ (C) and $d g_R/d\ln b$ (D).}
\label{TwoLoopDiags}
\end{figure}

The remaining two loop diagrams are shown in Fig.~\ref{TwoLoopDiags}.
Their evaluation in $d=6-\epsilon$ dimensions is now a straightforward exercise in combinatorics and standard momentum integrals, see, e.g., refs.~\cite{Alcantara81,Fei2015}. The resulting anomalous dimensions and the beta function are given by
\begin{subequations}
\label{epseta}
\begin{align}
\eta_\phi &= x_R\left(\frac{8}{3}-\frac{22\epsilon}{9}\right) - x_R^2\left(\frac{88}{9} - \frac{242\epsilon}{27} \right),\\
\eta_z &= x_R\left(\frac{4}{3}-\frac{11\epsilon}{9}\right) - x_R^2\left(\frac{176}{27} - \frac{484\epsilon}{81}\right),
\end{align}
\end{subequations}
and
\begin{equation}
\label{epsBeta}
\gdif{x_R}{\ln b} = \epsilon x_R - x_R^2\left(\frac{20}{3} - \frac{55\epsilon}{9}\right) + x_R^3\left(\frac{4160}{27} - \frac{1936\epsilon}{81}\right).
\end{equation}
Here, $\eta_{\phi/z} = dZ_{\phi/z}/d\ln b$ where $Z_{\phi/z}$ are the wavefunction renormalization parameters of the respective fields and $b$ corresponds to the momentum shell scaling factor in a Wilsonian RG picture~\cite{Herbut2007}. Furthermore, $x_R = g_R^2 = (g^2 Z_g^2  b^{6-d}) / (Z_\phi^2 Z_z) $ corresponds to the renormalized Yukawa coupling and thus $dx_R/d\ln b = 2g_R\cdot d g_R/d\ln b$. For conciseness, the standard  rescaling of $x_R$ with $S_d/(2\pi)^d$ has also been performed.

For small $\epsilon$, besides the Gaussian fixed point there are two other fixed points, one $\mathcal{O}(1)$ (``non perturbative") and the other $\mathcal{O}(\epsilon)$  (``perturbative"):
\begin{subequations}
\begin{align}
x_{*,\rm{nonp}} &= \frac{9}{208} - \frac{4947\epsilon}{27040} - \frac{186337 \epsilon ^2}{281216},\label{epsFPPat}\\
x_{*,\rm{pert}} &= \frac{3\epsilon}{20} + \frac{263\epsilon^2}{400},\label{epsFPPhy}
\end{align}
\end{subequations}
with
\begin{subequations}
\begin{align}
\theta_{\rm{nonp}} &= \frac{15}{52}-\frac{9421 \epsilon }{2704} -\frac{347939 \epsilon ^2}{105456},\\
\theta_{\rm{pert}} &= -\epsilon + \frac{52\epsilon^2}{15}.
\end{align}
\end{subequations}
Here, $\theta = \partial_{x_R} ( d x_R/d \ln b)_{x_R = x_*}$ denotes the universal exponent which determines the stability of the fixed point. The second fixed point~\eqref{epsFPPhy} is stable for small $\epsilon$ and corresponds to the one found in~\cite{OneLoop16}. The other one, $x_{*,\rm{nonp}}$, is unstable for small epsilon, and also not $\mathcal{O}(\epsilon)$, and it is therefore unclear whether it should be taken seriously within the present approach.
\begin{figure}[t]
\centering
\includegraphics[width=0.5\textwidth]{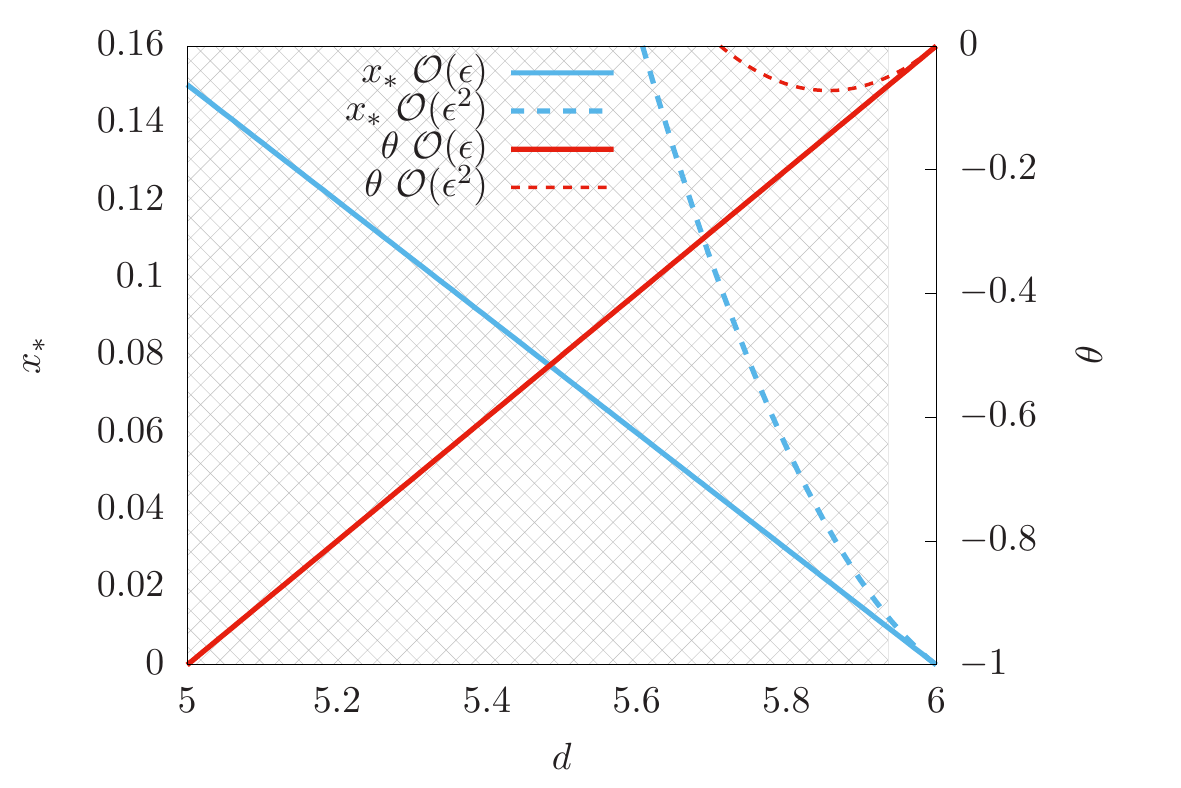}
\caption{$\epsilon$-expansion results for the values of the physical fixed point $x_{*,\m{pert}}$ and universal exponent $\theta$ as a function of spatial dimension. The solid lines provide the one-loop result whereas the dashed lines represent (na\"ively expanded) two-loop values. The shaded region indicates non-existence of the fixed point at two-loop level due to merging and annihilation (see main text).}
\label{EpsResult}
\end{figure}

It is now tempting to infer the value of the physical fixed point~\eqref{epsFPPhy} and its concomitant critical exponents, such as anomalous dimensions, 
\begin{equation}
\eta_{\phi,*} = \frac{2\epsilon}{5} + \frac{7\epsilon^2}{6} , \quad \eta_{z,*} = \frac{\epsilon}{5} + \frac{41\epsilon^2}{75},
\end{equation}
all the way to $\epsilon=1$, i.e. in the physical dimension $d=5$; see the dashed lines in Fig.~\ref{EpsResult}. This, however, appears to be of questionable validity. The $\epsilon$-expanded $\theta_{\m{pert}}$ becomes positive for $\epsilon > 15/52$, and therefore the perturbative fixed point loses its stability before $d=5$ is reached. Such a change of stability can only occur due to a collision with another fixed point. In fact, going back to the beta-function in eq. (4) for a more complete analysis and not expanding the $x_*$ about $\epsilon$, one finds that at a much lower value of  $\epsilon_c = \frac{60 \left(471-8 \sqrt{3298}\right)}{10769} \approx 0.064$, the fixed points $x_{*,\rm{pert}}$ and $x_{*,\rm{nonp}}$ had already collided. For all $\epsilon > \epsilon_c$, the non-trivial zeros of the beta function ~\eqref{epsBeta} become complex (shaded region in Fig.~\ref{EpsResult}).

Such a \emph{collision} and \emph{annihilation} of fixed points is a ubiquitous phenomenon, believed to be relevant in many different contexts; see e.g.~\cite{Kaveh2005,Kaplan2009, Herbut2014, Braun2014, Herbut2016}. Interestingly, if indeed occuring at some $\epsilon_c <1$, it would reconcile the results of~\cite{OneLoop16} with the expectation based on the arguments of ref.~\cite{FROHLICH1982}. The quadratic nature of the flow equation~\eqref{epsBeta} at the one-loop level does not allow a fixed-point collision for the simple reason that there is no other non-Gaussian fixed point to collide with. In a one-loop calculation, therefore, the stable non-Gaussian fixed point is bound  to exist all the way to $d=5$ (solid lines in Fig.~\ref{EpsResult}). Possible removal of it by collision and  annihilation with another fixed point could thus only be detected by going at least to the two-loop level.

Let us now discuss the reliability of our findings. The detected smallness of $\epsilon_c$ at two loops certainly lends some credibility to the perturbative result. On the other hand, it is not clear whether the \emph{non-existence} of the stable fixed point needs to persist all the way to $\epsilon=1$. As can be seen from the expressions~\eqref{epseta} and \eqref{epsBeta}, next-to-leading order corrections in the $\epsilon$-expansions are sizable and convergence properties of the $\epsilon$-expansion do not look encouraging. A higher order computation would certainly be desirable. Leaving this for future work, since the conclusion at the two-loop level is quite sensitive to the size of the (large) coefficients in the last terms in eqs. (3) and (4), in the next section we turn to a computation of the corresponding RG equations still at two loops, but directly in $d=5$.

\section{Fixed Dimension RG}
\label{SecFixDim}
In $d=5$ dimensions, $g$ is a canonically relevant coupling with mass dimension $[g] = \frac{1}{2}$. For situations like this, a renormalization scheme at fixed dimension was developed by Parisi~\cite{Parisi1998}. In this approach the bare mass of the fields is kept finite during the computation in order to keep infrared singularities under control. Renormalization is performed by means of the BHP formula~\cite{Parisi1998}
\begin{equation}
\label{BPHZ}
V_R(q) = V(q) - \sum_{n=0}^{\mathcal{D}}\frac{1}{n!}\partial_q^n V(q)\big|_{q=0}\cdot q^n
\end{equation}
Here, $V_{(R)}(q)$ is the (renormalized) value of some UV-divergent diagram at external momentum $q$ and $\mathcal{D}$ is the highest degree of divergence involved. If the divergence of some diagram is due to substructures, the latter themselves have to be renormalized first~\cite{Parisi1998}.

To obtain the flow equations in this scheme, we need to re-evaluate the diagrams in Figs.~\ref{OneLoopDiags} and~\ref{TwoLoopDiags}. While the combinatoric factors and tensor algebra contributions do not change, the momentum integrals have to be computed anew. Doing so numerically and rescaling $x_R$ to fit the results of the $\epsilon$-expansion at one-loop level yields
\begin{subequations}
\label{Fixeta}
\begin{align}
\eta_\phi &= \frac{8}{3}x_R + 12.5103 x_R^2,\\
\eta_z &= \frac{4}{3}x_R + 8.3402 x_R^2,
\end{align}
\end{subequations}
and
\begin{equation}
\label{FixBeta}
\frac{dx_R}{d\ln b} = x_R - \frac{20}{3}x_R^2 + 185.405 x_R^3.
\end{equation}
Eq.~\eqref{FixBeta} has only complex-valued fixed-point solutions, aside from $x_*=0$. This would be in accord with the conclusion that the stable, real,  non-trivial fixed point has disappeared on its descend from six dimensions.

\begin{figure}[t]
\centering
\includegraphics[width=0.5\textwidth]{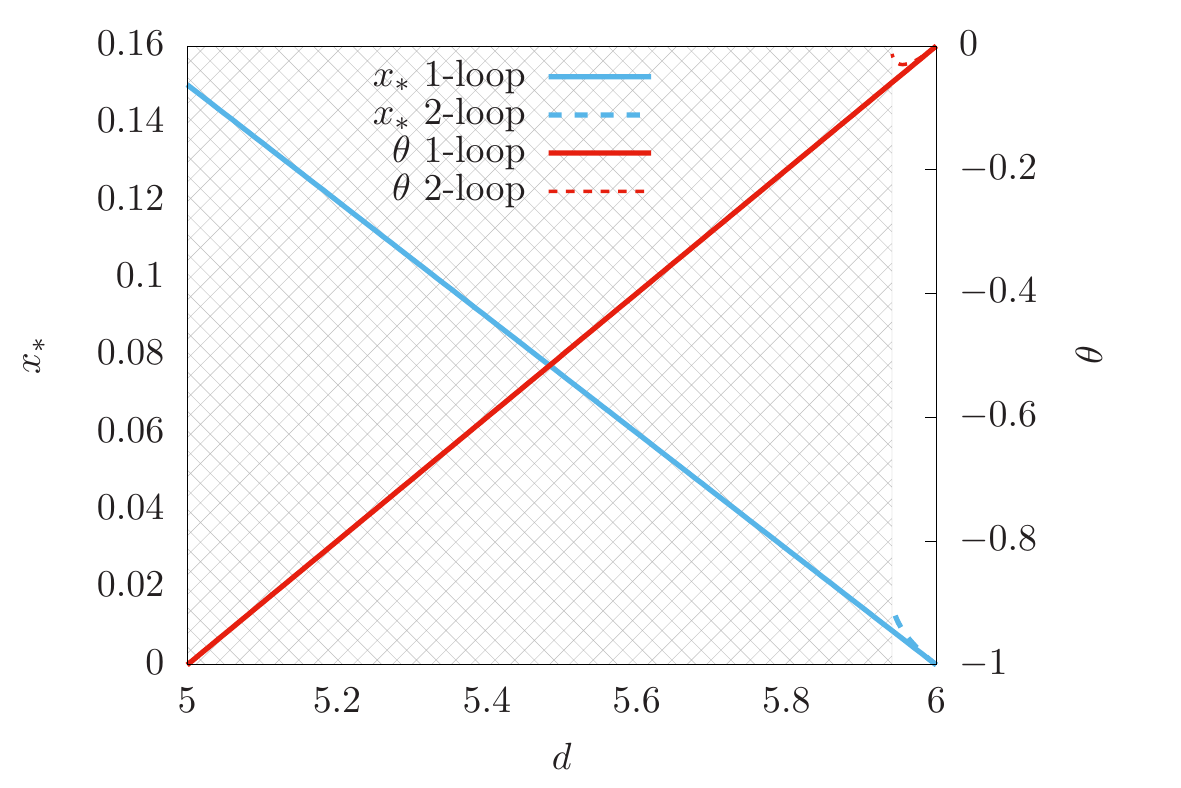}
\caption{Fixed-dimension RG results for the fixed-point values and universal exponent $\theta$ at one (solid lines) and two (dashed lines) loop level. The values for $d>5$ are generated by an artificial extension of the formalism, see main text. Good agreement with results from the $\epsilon$-expansion is observed, cf. Fig.~\ref{EpsResult}.}
\label{FixedResult}
\end{figure}

As an additional check, let us pretend for a moment that the coefficients in eqns.~\eqref{Fixeta} and~\eqref{FixBeta} were independent of dimension and reinstate the general factor $(6-d)$ to the linear term in the RG equation~\eqref{FixBeta} for $x_R$. The calculation then can again be continued in  dimensionality, in this case towards six dimensions. The corresponding results are shown in Fig.~\ref{FixedResult}.
Very similar to our findings in Sec.~\ref{SecEps}, we recover a real, stable fixed point solution in close vicinity to six spatial dimensions. Quantitatively, this would correspond to an $\epsilon_c = 0.045$ in decent agreement with the value from the $\epsilon$-expansion. Again, the extension of fixed-dimension results should not be taken as a reliable quantitative result. However, it is reassuring to see that its findings agree with the results from the $\epsilon$-expansion.

\section{Functional RG}
\label{SecfRG}
Aside from the question of convergence in terms of $\epsilon$, there is also a reason to be concerned about the perturbative expansion itself, affecting the results of both Secs.~\ref{SecEps} and~\ref{SecFixDim}. In both cases, the coefficients of the cubic terms in the flow equation for $x_R$ are more than a magnitude larger than those of the quadratic terms. It is therefore not safe to assume that higher order contributions would yield only small corrections. On the contrary, it can be expected that rather high order calculations and elaborate resummation schemes may be necessary to achieve a satisfactory level of reliability (see, e.g.,~\cite{Gracey2015}).

In this work, we do not attempt to pursue this direction. Instead, we employ an inherently non-perturbative method, the functional Renormalization Group (fRG). Recently, the scalar extension~\cite{Fei2014} of the $O(N)$ model has been investigated successfully~\cite{Eichhorn2016} by means of this approach. It is thus only natural to apply the method to the tensorial theory as well. The central component of fRG is the exact Wetterich equation~\cite{Wetterich1993} for the effective average action $\Gamma_k$,
\begin{equation}
\label{Wettereq}
\partial_k \Gamma_k = \frac{1}{2}\Tr\left[\frac{\partial_k R_k}{\Gamma_k^{(2)} + R_k}\right].
\end{equation}
Here, $k$ is the scale parameter that is used instead of $\ln b$ for conciseness. $\Gamma_k$ itself interpolates between the microscopic action $S$ for $k\rightarrow\Lambda$ with $\Lambda$ the overall UV cutoff and the full effective action $\Gamma$ in the deep infrared, $k\rightarrow 0$. This is achieved by means of the regulator function $R_k$, which has to fulfill certain properties~\cite{Wetterich1993}, such as to vanish for $k\rightarrow 0$.

The right hand side of eq.~\eqref{Wettereq} is essentially the trace of the (regulated) propagator matrix $(\Gamma_k^{(2)})^{-1}$, so the fundamental structure of the equation is strictly one-loop. Since $\Gamma_k^{(2)}$ is the \emph{full} inverse propagator, higher loop contributions are implicitly accounted for and the equation remains in principle exact. For practical purposes, however, approximations have to be made to be able to solve eq.~\eqref{Wettereq}. One such approximation would be to replace $\Gamma_k$ in the right hand side of the equation by the microscopic action $S$. In this case, the Wetterich equation can be integrated explicitly~\cite{Berges2002}, yielding
\begin{equation}
\label{Wetterone}
\Gamma^{\m{1-loop}} = S + \frac{1}{2}\Tr\ln S^{(2)}.
\end{equation}
This is the equation for the perturbative one-loop effective action. It can thus be inferred that the Wetterich equation reproduces perturbative one-loop renormalization group as a limiting case.

Moving beyond the one-loop level, comparison to perturbation theory becomes far less obvious~\cite{WetterBrock95}. It is clear that simply setting $\Gamma_k$ equal to the microscopic action supplemented with running couplings cannot account for this anymore as it yields precisely the one-loop limit. Therefore, a more elaborate \emph{truncation} has to be used as an ansatz for $\Gamma_k$ in eq.~\eqref{Wettereq}.

The full quantum effective action $\Gamma$ will generally consist of all, even canonically irrelevant, operators compatible with the symmetries of the system. Furthermore,  frequency and momentum dependence of those $n$-point functions have to be taken into account as well. Therefore, we have two ways of systematically expanding our truncation in order to encompass the two-loop calculations from the sections above.

\paragraph{Quartic operators}
Let us first consider diagram D in Fig.~\ref{TwoLoopDiags}. With only the Yukawa coupling term available, this diagram cannot be composed by combining one-loop subdiagrams. Consequently, the only possible way to achieve a vertex renormalization at all in our one-loop fRG scheme must be to include higher order operators in our ansatz for $\Gamma_k$. It could be argued that such operators are canonically irrelevant and should therefore be negligible in the deep infrared. However, this does not preclude them from crucially affecting the RG flow of $x_R$ at intermediate scales. This is particularly true for the \emph{functional} RG equations which are constructed differently than perturbative ones due to the one-loop structure of the Wetterich eq.~\eqref{Wettereq}.

\begin{figure}[t]
\centering
\includegraphics[width=0.4\textwidth]{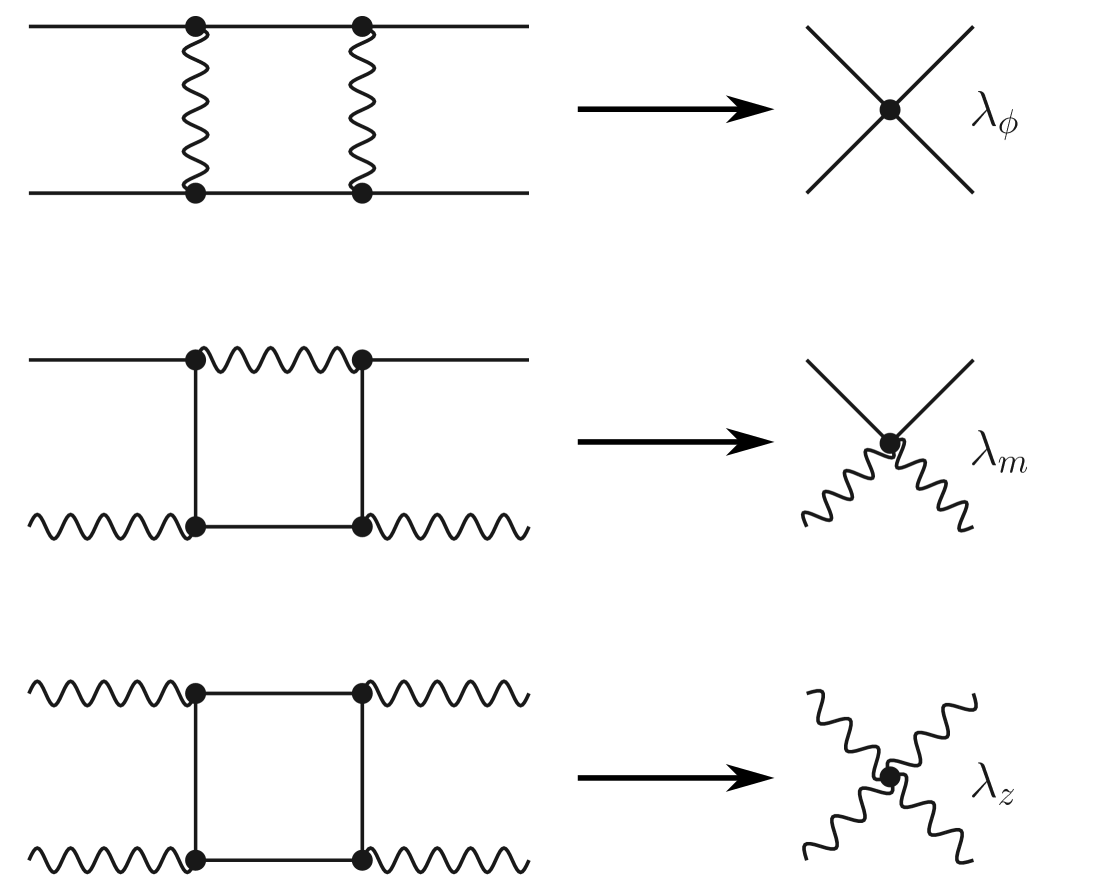}
\caption{Schematics for generating the quartic terms during the fRG flow by box type diagrams.}
\label{Boxer}
\end{figure}

The next highest order operator to be generated from the Yukawa term is $\lambda_{\phi,k} (\phi_i\phi_i)^2$. This is hardly surprising, since $g \phi_i \sigma^a_{ij}\phi_j z^a$ was supposed to result from a Hubbard-Stratonovich decoupling of this term in the first place. However, two more quartic operators are generated alongside, $\lambda_{m,k} \phi^2 z^a z^a$ and $\lambda_{z,k}(z^a z^a)^2$, see Fig.~\ref{Boxer}.

\begin{figure}[t]
\centering
\includegraphics[width=0.5\textwidth]{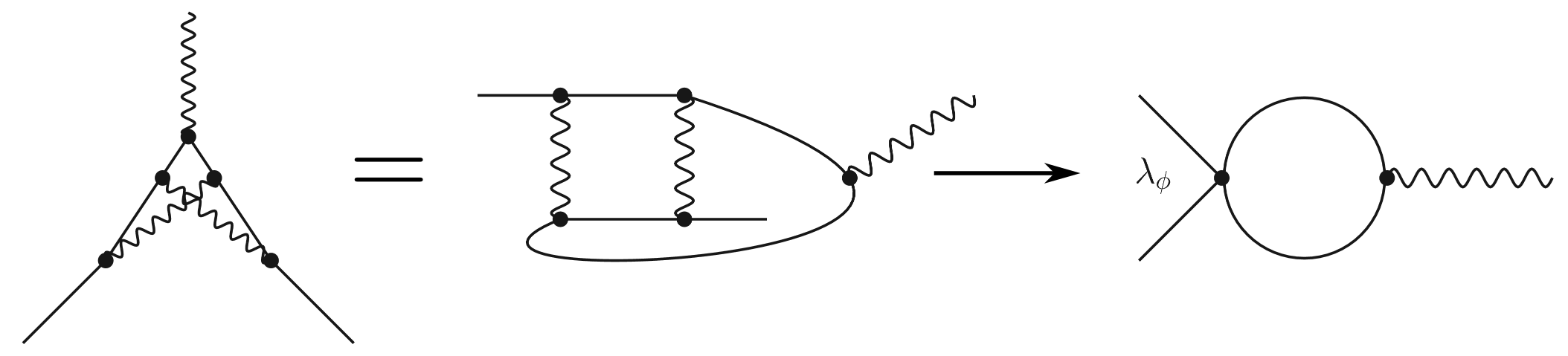}
\caption{Diagrammatic scheme: feedback of the running $\lambda_\phi$ coupling into the flow of $x_R$ accounts for the two-loop correction of the Yukawa vertex.}
\label{Tinnitus}
\end{figure}

As can be inferred from Fig.~\ref{Tinnitus}, the feedback of the flow of $\lambda_\phi$ does indeed account for the diagram structure D in Fig.~\ref{TwoLoopDiags} in the flow of $x_R$. In terms of diagram structures, the two loop perturbative expansion is thus recovered by the inclusion of $\lambda_\phi$. For consistency, however, $\lambda_m$ and $\lambda_z$ as well as contributions $\sim\lambda^2$ in the flow of the quartic couplings themselves must be accounted for as well in this scheme (see below). Therefore, the present fRG approach clearly goes beyond second order perturbation theory in this respect.

\paragraph{Momentum dependent vertices}
Secondly, the frequency and momentum dependence of vertices or, more generally, $n$-point functions must be taken into account to fully encompass two loop perturbation theory. This becomes obvious when comparing, for example, the middle and right diagrams in Fig.~\ref{Tinnitus}. The box-like subdiagram, which is provided by the $\lambda_{\phi,k}$ vertex in fRG, possesses a non-trivial momentum structure even if all external momenta vanish. A constant quartic coupling $\lambda_{\phi,k}$ cannot fully account for this. It would thus have to be made a function of frequency and momentum to fully encompass all effects which are included by construction in the perturbative expansion.

In this work, we refrain from pursuing this direction any further. While it is in principle possible to construct fRG truncations that do include such dependencies~\cite{WetterBrock95,Blaizot2006,Benitez2009,Benitez2012,Rose2018}, their application and evaluation is often a formidable task and can generally only be performed numerically. Here, we want to focus on aspects that can be investigated at least semi-analytically.

Summarizing, our ansatz for the effective average action is given by
\begin{equation}
\label{fRGAnsatz}
\begin{aligned}
\Gamma_k =& \int d^d{x}\left[\frac{1}{2}\phi_i[m_{\phi,k}^2-Z_\phi\partial_\mu^2]\phi_i + \frac{1}{2}z_a[m_{z,k}^2-Z_z\partial_\mu^2] z_a \right.\\
&+ g_kz_a\phi_i\sigma^a_{ij}\phi_j + \lambda_{\phi,k} (\phi_i\phi_i)^2\\
&+ \lambda_{m,k} z_a z_a\phi_i\phi_i + \lambda_{z,k} (z_a z_a)^2  \Bigg].
\end{aligned}
\end{equation}
Note that we do include the running masses $m_{\phi,k}^2$ and $m_{z,k}^2$ in our truncation. It is generally impossible to consistently tune dimensionless fixed point mass parameters to zero away from the gaussian fixed point.

By employing spatio-temporally constant coupling parameters, we do not encompass second order perturbation theory completely. On the other hand, diagrammatic structures are accounted for even beyond the perturbative two-loop level already in this setup.

For our concrete evaluations, we choose the optimized Litim cutoff function~\cite{LitimPLB2000,LitimPRD2001}
\begin{equation}
R_{k,\phi/z}(q) \equiv k^2r_{k,\phi/z}(q) = Z_{\phi/z}\left(k^2-q^2\right)\Theta\left[k^2-q^2\right],
\end{equation}
where $\Theta[\cdot]$ is the Heaviside step function.
Given eq.~\eqref{fRGAnsatz}, the flow equations for $x_R$ and the dimensionless couplings
\begin{subequations}
\label{DimlessQuart}
\begin{align}
&\qquad\qquad m_\phi^2 = \frac{m_{\phi,k}^2}{Z_\phi k^2}, \qquad m_z^2 = \frac{m_{z,k}^2}{Z_z k^2}\\
&\lambda_\phi = \frac{\lambda_{\phi,k}k^{d-4}}{Z_\phi^2},\quad \lambda_m = \frac{\lambda_{m,k}k^{d-4}}{Z_\phi Z_z},\quad \lambda_z = \frac{\lambda_{z,k}k^{d-4}}{Z_z^2},
\end{align}
\end{subequations}
may then readily be computed (see App.~\ref{SecFRGApp} for details). After proper rescaling, we find
\begin{subequations}
\label{FRGeta}
\begin{align}
\eta_\phi &= \frac{8}{3}\frac{x_R}{(1+m_\phi^2)^2(1+m_z^2)^2},\\
\eta_z &= \frac{4}{3}\frac{x_R}{(1+m_\phi^2)^4},\\
\end{align}
\end{subequations}

\begin{subequations}
\label{FRGmass}
\begin{align}
\frac{d m_\phi^2}{d\ln k} &= (2 - \eta_\phi)m_\phi^2 + 4x_Rc_{1,1}^d - 32\lambda_\phi c_{1,0}^d - 8\lambda_m c_{0,1}^d\\
\frac{dm_z^2}{d\ln k} &= (2 - \eta_z)m_z^2 + 2x_R c_{2,0}^d - 8\lambda_m c_{1,0}^d - 32\lambda_z c_{0,1}^d
\end{align}
\end{subequations}

\begin{equation}
\label{FRGBeta}
\frac{dx_R}{d\ln k} = \left[(6-d) - 2\eta_\phi - \eta_z\right]x_R + 32x_R \lambda_\phi c_{2,0}^d + 32x_R\lambda_m c_{1,1}^d,
\end{equation}
and
\begin{subequations}
\label{FRGlambda}
\begin{align}
\frac{d\lambda_\phi}{d\ln k} =& \left[(4-d) - 2\eta_\phi\right]\lambda_\phi + x_R^2 c_{2,2}^d + 16x_R\lambda_\phi c_{2,1}^d \nonumber\\
& - 4x_R\lambda_m c_{1,2}^d + 80\lambda_\phi^2 c_{2,0}^d + 4\lambda_m^2 c_{0,2}^d\\
\frac{d\lambda_m}{d\ln k} =& \left[(4-d) - \eta_\phi - \eta_z\right]\lambda_m + 2x_R^2 c_{3,1}^d - 16x_R\lambda_\phi c_{3,0}^d \nonumber\\
& - 12x_R\lambda_m c_{2,1}^d - 16x_R\lambda_z c_{1,2}^d + 32\lambda_{\phi}\lambda_m c_{2,0}^d \nonumber\\
& + 16\lambda_m^2 c_{1,1}^d + 32\lambda_m\lambda_z c_{0,2}^d\\
\frac{d\lambda_z}{d\ln k} =& \left[(4-d) - 2\eta_z\right]\lambda_z + \frac{1}{2}x_R^2 c_{4,0}^d - 4x_R\lambda_m c_{3,0}^d \nonumber\\
& + 4\lambda_m^2 c_{2,0}^d + 80\lambda_z^2 c_{0,2}^d,
\end{align}
\end{subequations}
where  $c_{i,j}^d$ is given in eq.~\eqref{Sreschold} (see App.~\ref{SecFRGApp} for details). $x_R$ was rescaled in such a way that the perturbative one-loop equations are recovered when setting $\lambda_{\phi/m/z}$ and $m^2_{\phi/z}$ to zero and ignoring the $\eta_{\phi/z}$-dependence inside of the $c_{i,j}^d$.

\begin{figure}[t]
\centering
\includegraphics[width=0.5\textwidth]{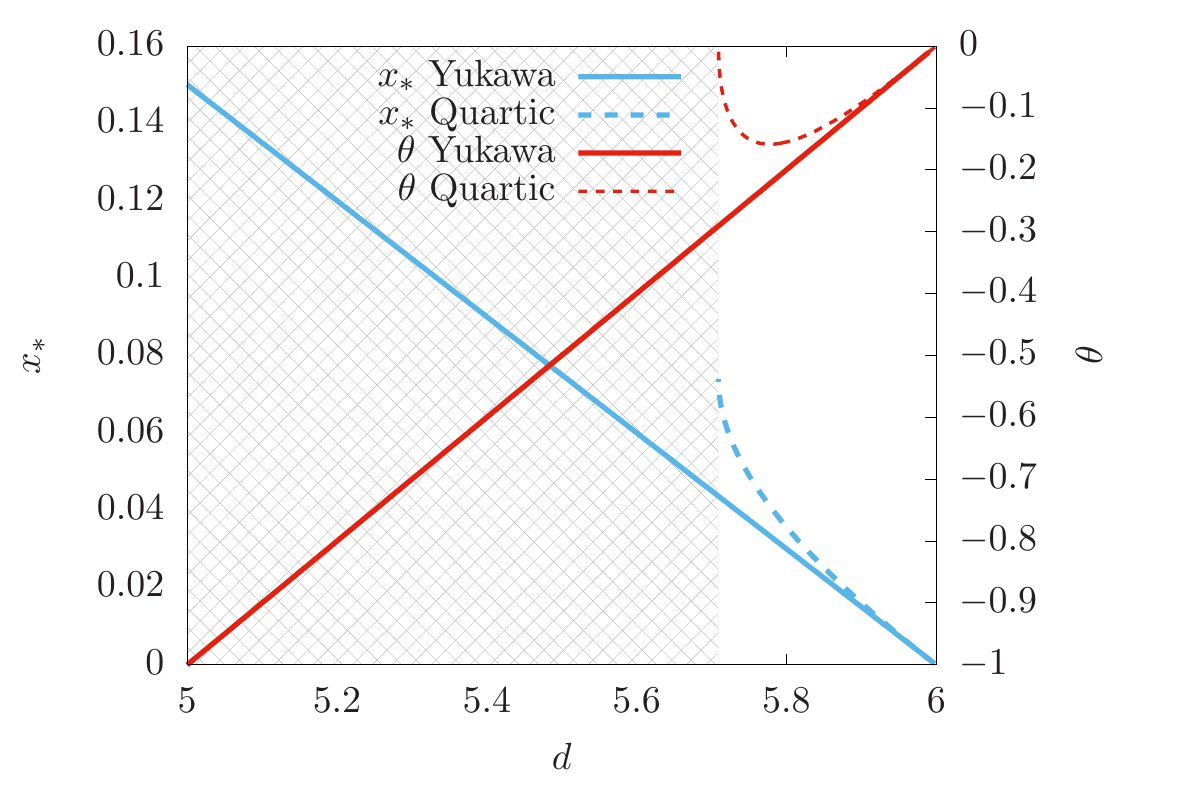}
\caption{Functional RG results for the fixed-point values and universal exponent $\theta$ at one loop level (solid lines) and including quartic terms (dashed lines). Non-existence of an entirely stable fixed point due to annihilation is indicated by shading.}
\label{fRGResult}
\end{figure}

The solution to these RG equations generally yields a multitude of fixed points. There is, however, at most one completely stable solution (except for the flow of the mass parameters which is always relevant). This fixed point exists in the vicinity of $d=6$ dimensions and is annihilated at $d = 5.71$ (corresponding to $\epsilon_c = 0.29$) after merging with another, unstable solution, see Fig.~\ref{fRGResult}. While this value for the limiting $\epsilon$ is thus somewhat larger than in the perturbative analyses, the general finding prevails. Incidentally, the numerical value of $\epsilon_c$ found in the fRG is very close to the value where the stability exponent in eq. (6.b) changes sign.

The fRG analysis we performed is by no means exact and fails to encompass perturbation theory in some respects, but it also goes beyond perturbative RG in others. The stability of the result with respect to such alterations may give some confidence that the non-existence of an IR-stable fixed point for the tensorial $O(2)$ model holds also beyond second-order perturbation theory.

\section{Conclusions}
\label{SecConc}
In this work, we revisited the $O(2)$ theory in  tensorial representation near six dimensions. Previous analysis at one-loop level found a real, infrared-stable,  non-Gaussian fixed-point. Although, as discussed in the introduction, not directly violating them, this finding seems to go against the intuition based on  proofs~\cite{Aizenman81,FROHLICH1982} that state that at least for $N=1$ and $N=2$  the standard $\phi^4$-theory must be trivial for any $d>4$. Motivated by this observation, we extended our perturbative analysis of the $O(2)$ model to the two-loop level. Annihilation of the stable fixed-point solution at $d= 5.94$ was found to occur in the $\epsilon$-expansion at this order. While clearly this conclusion may change, if surviving higher-order or more elaborate computations it would point to triviality in $d=5$. The convergence properties of the $\epsilon$-expansion, however, seem problematic. In order to check the robustness of this result with respect to the employed RG scheme, we performed the same analysis within Parisi's fixed-dimension RG  directly in $d=5$, where again no real, non-trivial fixed point could be found, further strengthening this conclusion.

Finally, we investigated the same model by means of functional renormalization group. We were able to show that within our truncation, effects of higher loop orders are partially included as well. At the same time, the treatment of internal momentum structures of multi-loop diagrams is not equivalent to the perturbative situation anymore. Nevertheless, we again found that a stable fixed point only exists in the immediate vicinity of six dimensions.

It is of course possible that a more elaborate analysis might reveal yet another stable fixed point, or reinstate the one which appears to be annihilated in our treatment. While it goes beyond the scope of our present work, it would therefore be worthwhile to compute third and higher-loop corrections and employ resummation techniques as well. Furthermore, in ref.~\cite{Eichhorn2016} on the scalar extension of the $O(N)$ model, higher order truncations and even an analytic investigation of the fixed-point effective potential were employed. While these did not yield qualitatively different results from a low order analysis such as ours, it would still be interesting to pursue a similar program for the tensorial extension.

Last but not least, it is important to extend the higher-order analysis also beyond the $N=2$ case. Even if the results of the present work prove to persist to higher loops, it is conceivable that stable fixed points could survive in $d=5$ for $N>2$.

\begin{acknowledgments}
We are grateful to John Gracey, Holger Gies, Michael Scherer, and Igor Boettcher for enlightening discussions and correspondence. We thank Igor Boettcher, Lukas Janssen and Michael Scherer for useful comments on the manuscript. The authors acknowledge support by the NSERC of Canada. D.R. is supported, in part, by the German Research Foundation (DFG) through the Institutional Strategy of the University of Cologne within the German Excellence Initiative (ZUK 81).
\end{acknowledgments}


\appendix

\section{Derivation of the fRG flow equations}
\label{SecFRGApp}
Here, we provide some details on how to derive the flow equations~\eqref{FRGeta}-\eqref{FRGlambda} from the ansatz~\eqref{fRGAnsatz}. First, the exact RG equation~\eqref{Wettereq} itself can be expanded as
\begin{equation}
\label{PmoneF}
\partial_t \Gamma_k = \frac{1}{2}\m{Tr}\left[\tilde{\partial}_t \ln\left(\mathcal{P}_k\right) - \tilde{\partial}_t\sum_{n=1}^\infty\frac{(-1)^n}{n}\left(\mathcal{P}_k^{-1}\mathcal{F}_k\right)^n\right],
\end{equation}
where $\tilde{\partial}_t$ denotes the scale derivative $\partial_{\ln k}$ acting exclusively on the regulator dependence of any term it is applied to. Furthermore, the full propagator $\Gamma_k^{(2)}$ is split into field-dependent ($\mathcal{F}_k$) and inverse field-independent ($\mathcal{P}_k^{-1}$) parts. Projection rules onto the flow of the respective running couplings can now be constructed accordingly:
\begin{widetext}
\begin{subequations}
\begin{align}
\eta_\phi\phi_i\phi_i\Omega &= -\frac{\partial_{\ln k} Z_\phi}{Z_\phi}\phi_i\phi_i\Omega = \frac{1}{4Z_\phi}\partial_q^2\tilde{\partial}_t\m{Tr}\left[\mathcal{P}_k^{-1}(p)\mathcal{F}_k \mathcal{P}_k^{-1}(p-q)\mathcal{F}_k\right]_{\phi^2,q = 0}\\
\eta_z z_a z_a\Omega &= -\frac{\partial_{\ln k} Z_z}{Z_z}z_a z_a\Omega = \frac{1}{4Z_z}\partial_q^2\tilde{\partial_t}\m{Tr}\left[\mathcal{P}_k^{-1}(p)\mathcal{F}_k \mathcal{P}_k^{-1}(p-q)\mathcal{F}_k\right]_{z^2,q=0}\\
\partial_{\ln k} g_k z_a\phi_i\sigma^a_{ij}\phi_j \Omega &= -\frac{1}{2}\tilde{\partial}_t\m{Tr}\left[\frac{1}{2}\left(\mathcal{P}_k^{-1}\mathcal{F}_k\right)^2 - \frac{1}{3}\left(\mathcal{P}_k^{-1}\mathcal{F}_k\right)^3\right]_{z\phi^2}\\
\partial_{\ln k}\lambda_{\phi,k}(\phi_i\phi_i)^2\Omega &= -\frac{1}{2}\tilde{\partial}_t\m{Tr}\left[\frac{1}{2}\left(\mathcal{P}_k^{-1}\mathcal{F}_k\right)^2 - \frac{1}{3}\left(\mathcal{P}_k^{-1}\mathcal{F}_k\right)^3 + \frac{1}{4}\left(\mathcal{P}_k^{-1}\mathcal{F}_k\right)^4\right]_{\phi^4}\\
\partial_{\ln k}\lambda_{m,k} z_a z_a\phi_i\phi_i\Omega &= -\frac{1}{2}\tilde{\partial}_t\m{Tr}\left[\frac{1}{2}\left(\mathcal{P}_k^{-1}\mathcal{F}_k\right)^2 - \frac{1}{3}\left(\mathcal{P}_k^{-1}\mathcal{F}_k\right)^3 + \frac{1}{4}\left(\mathcal{P}_k^{-1}\mathcal{F}_k\right)^4\right]_{\phi^2z^2}\\
\partial_{\ln k}\lambda_{z,k} \m{Tr}\left[\left(z_a\sigma^a\right)^4\right]\Omega &= -\frac{1}{2}\tilde{\partial}_t\m{Tr}\left[\frac{1}{2}\left(\mathcal{P}_k^{-1}\mathcal{F}_k\right)^2 - \frac{1}{3}\left(\mathcal{P}_k^{-1}\mathcal{F}_k\right)^3 + \frac{1}{4}\left(\mathcal{P}_k^{-1}\mathcal{F}_k\right)^4\right]_{z^4},
\end{align}
\end{subequations}
\end{widetext}
where all fields have been set to $\phi_i(p) = \phi_i\delta(p)$ or $z_a(p) = z_a\delta(p)$, respectively and $\Omega$ is the spacetime volume.

Computing the traces and comparing coefficients is now straightforward algebra which can be performed without even choosing a specific shape of the regulator function. In order to compute the momentum integrals, however, $R_k$ needs to be specified. Two standard integrals occur, whose values we provide for the optimized cutoff~\cite{LitimPLB2000,LitimPRD2001} used in this work:
\begin{widetext}
\begin{equation}
I_1(n_\phi,n_z,d) = \tilde{\partial}_t \int_p\frac{1}{Z_\phi^{n_\phi}\left[p^2 + \tilde{m}_{\phi,k}^2 + k^2 r_k\right]^{n_\phi}}\frac{1}{Z_z^{n_z}\left[p^2 + \tilde{m}_{z,k}^2 + k^2 r_k\right]^{n_z}} = \frac{2 c_{n_\phi,n_z}^d S_d k^{d-2n_\phi-2n_z}}{(2\pi)^dZ_\phi^{n_\phi} Z_z^{n_z}\left[1+m_\phi^2\right]^{n_\phi}\left[1+m_z^2\right]^{n_z}}
\end{equation}
with
\begin{equation}
\label{Sreschold}
c_{i,j}^d = \frac{i\eta_\phi}{d(d+2)\left[1+m_\phi^2\right]} + \frac{j\eta_z}{d(d+2)\left[1+m_z^2\right]}- \frac{i}{d\left[1+m_\phi^2\right]} - \frac{j}{d\left[1+m_z^2\right]}
\end{equation}
and
\begin{equation}
\begin{aligned}
I_2(n_\phi,n_z,d) &= \partial_q^2 \tilde{\partial}_t \int_p\frac{1}{Z_\phi^{n_\phi} Z_z^{n_z}\left[p^2 + m_{k,\phi/z}^2 + k^2 r_k(p)\right]\left[(p-q)^2 + m_{k,\phi/z}^2 + k^2 r_k(p-q)\right]}\bigg|_{q=0}\\
&= \frac{8k^{d-6}}{3Z_\phi^{n_\phi} Z_z^{n_z}}\frac{S_d}{(2\pi)^d}\frac{1}{[1+m_\phi^2]^{2n_\phi}[1+m_z^2]^{2n_z}}
\end{aligned}
\end{equation}
For the last equation, it is implicitly assumed that $n_\phi + n_z = 2$ and the mass parameters in the propagators have to be chosen accordingly.
\end{widetext}

\bibliography{Looper}

\end{document}